# Complemetarity of Phases in Aharonov-Bohm Solenoid Effect

Y.Ben-Aryeh

Physics Department, Technion-Israel Institute of Technology, Haifa, 32000 Israel

e-mail: phr65yb@ph.technion.ac.il

In the present Note it is suggested that there should be a certain complementarity of phases between Aharonov-Bohm (AB) solenoid phase calculation on one part of the system and a phase calculation about another part of the physical system. Assuming a unique value for the function of the total system, after a complete circulation of the electron around the solenoid, the sum of these two phase changes should vanish. Such assumption leads to a compatibility relation between our previous calculations for the AB solenoid phase effect and that of the original calculation by AB.

**1.Introduction**

The explanation of Aharonov-Bohm (AB) solenoid effect made in 1959 has been quite 'revolutionary' [1]. While classically the equations of motion are described by classical fields, in Quantum Mechanics (QM) the potentials are the essential parameters. Although in the AB solenoid effect there is no force on the electron the quantum phase is obtained by taking into account the vector potential [1]. This effect has been explained further by assuming that the potentials leads to 'modular variables' which show the physical features of QM non-locality [2,3].

In a previous paper [4] we claimed that during the motion of the electron around the solenoid a force is produced on the solenoid equal in magnitude and opposite in sign to the rate of change of the momentum of the em field. We have shown that the integral of energy of interference over the corresponding time, gives the AB effect, up to a possible change in the phase sign. Since I have no doubt about the correctness of the mathematical calculation in our previous work [4] the question arises how this result is compatible with AB explanation. I suggest in the present Note that there should be a certain complementarity of phases between AB calculation on one part of the system and a phase calculation about another part of the physical system. I explain further this idea in the next section.

**2. Complementarity of phases calculations**

It has been shown in previous papers that the AB effect is a topological effect where its explanation should be made in the context of other topological effects [5-8]. There are two alternative approaches for describing geometric phases:

The conventional approach to 'parallelism' in Schrodinger equation is given by

$$\langle \Phi | d\Phi \rangle = 0 \qquad . \qquad (1)$$

This condition is obtained in Schrodinger equation by requiring that $|\Phi(t+\Delta t)\rangle$ is in phase with $|\Phi(t)\rangle$, which means that $\langle \Phi(t) | \Phi(t+\Delta t) \rangle$ is real and positive. Since $\langle \Phi(t) | \Phi(t) \rangle = 1$ and $\langle \Phi(t) | d\Phi(t) \rangle$ is necessarily imaginary due to normalization, then the assumption made in Eq. (1) follows from the use of 'parallelism' in Schrodinger equation. The connection of Eq.(1) is similar to the connection used by Pancharatnam quite long ago [9,7], who defined two light beams to be in phase if the intensity of their sum is maximal. The basic idea is that phases do not follow the transitive law so that if system $A$ is in phase with system $B$ and system $B$ is in phase with system $C$, in general cases it does not follow that system $A$ is in phase with system $C$. This idea can be applied to the connection of Eq.(1) as *locally* the phase change is vanishing but *globally* on a closed circuit it leads to the geometric phase contribution. Pancharatnam connection which is analogous to Eq.(1) has been used also for calculating geometric phases.

In the alternative vector potential approach the wavefunction $|\Psi(\vec{x},t)\rangle$ of Schrodinger equation depends on set of parameters $\vec{R}$ e.g. $(R_1, R_2, R_3)$ which are functions of time so that $|\Psi(\vec{x},t)\rangle = |\Psi(\vec{R}(t))\rangle$ then the geometric phase $F(C)$ can be calculated by

$$F(C) = \frac{1}{i} \oint_C \langle \Psi | \vec{\nabla}_R \Psi \rangle \cdot d\vec{R} \qquad . \tag{2}$$

One should notice that in the vector potential approach 'parallelism' in the Schrodinger equation can be defined *in an analogy to AB effect* by

$$d\phi_G = \langle \Psi | \vec{\nabla}_R \Psi \rangle \cdot d\vec{R} \qquad . \tag{3}$$

where $d\phi_G$ is the change in the geometric phase following parameters change $d\vec{R}$. One should notice, however, that the differential of Eq.(3) is arbitrary up to a gauge and only for a closed circuit $C$ we get a well defined geometric phase independent of gauge transformation.

The following example taken from Ref. [9] illuminates the difference between the two approaches. Let us assume a Hamiltonian in a two-dimensional space of the form

$$H = \begin{pmatrix} x & y \\ y & -x \end{pmatrix} = r \begin{pmatrix} \cos\phi & \sin\phi \\ \sin\phi & -\cos\phi \end{pmatrix} \qquad . \tag{4}$$

The eigenvalues are $\pm r$, with orthonormal eigenfunctions

$$|\chi_+\rangle = \begin{pmatrix} \cos\frac{\phi}{2} \\ \sin\frac{\phi}{2} \end{pmatrix} \quad , \quad |\chi_-\rangle = \begin{pmatrix} -\sin\frac{\phi}{2} \\ \cos\frac{\phi}{2} \end{pmatrix} \qquad . \tag{5}$$

Under such conditions

$$|\delta\chi_+\rangle = \frac{1}{2}|\chi_-\rangle\delta\phi \quad ; \quad |\delta\chi_-\rangle = -\frac{1}{2}|\chi_+\rangle\delta\phi \qquad , \tag{6}$$

so that

$$\langle \vec{\chi}_\pm | \delta\vec{\chi}_\mp \rangle = \langle \vec{\chi}_\pm | \nabla_\phi \vec{\chi}_\pm \rangle \cdot \delta\phi = 0 \qquad , \tag{7}$$

and Eq. (7), similar to Eq. (1), is satisfied. We can manipulate the phase factor in such a way that locally the connection of Eq.(1) is satisfied. However, as the phases *are not added transitively* there is a global phase change under a closed circuit. In the above approach the eigenkets *are not single-valued*, however, undergoing a change of sign when $\phi$ makes a full circuit from 0 to $2\pi$. It is quite easy to find in this case a phase factor that will make the kets *single-valued*. For $|\chi_-\rangle$ for example make the transformation

$$|X_-\rangle = |\chi_-\rangle \exp(i\phi/2) \qquad . \tag{8}$$

This transformation restores the *single-valuedness*, but now there is a non-zero vector potential

$$A_\phi = \frac{1}{i}\left\langle X_- \mid \frac{d}{d\phi} X_- \right\rangle = \frac{1}{2} \qquad , \tag{9}$$

so that $A_\phi d\phi$ does not vanish. This example shows that for certain calculations of geometric phases one has to choose between the approach of making the vector potential orthogonal to the movement $d\vec{R}$ leading to the connection

$$\vec{A}(\vec{R}) \cdot d\vec{R} = 0 \qquad , \tag{10}$$

and the approach of *single-valuedness* of the total wavefunction, leading to calculations with vector potentials for which $\vec{A}(\vec{R}) \cdot d\vec{R} \neq 0$, *like that made in AB effect*.

The crucial point in the above arguments is that the calculations with the vector potentials take care of the requirement that the *total wavefunction* will be *single* valued. One should notice that our calculation for the AB solenoid phase changes has been made for a certain part of the physical system ,i.e., the e.m. energy of interference which is different from that of the electron part made by the vector potential calculations. The results obtained in our previous work seem to be consistent with AB phase calculations only if we claim and that the sum of phases given by AB effect plus that made by us in [4] is vanishing due to the assumption that the *total* wavefunction is single valued.

The above 'conjecture' might be objected due to the fact that there has been given evidence for AB effect with magnetic field completely shielded from the electron wave [11,12]. However, the comlementarity of phases is not limited to a calculation of energy of interference. I guess that under a shielding of the solenoid there will be forces between the electron and the metallic shielding (although there is no force on the electron) so that such forces will yield phase changes which will be complementary to AB phase effects.

The problem of single valuedness for the total wavefunction plays important role in the definitions of bosons and fermions. Leinaas and Myrheim [13] raised the idea that the two possibilities of *identical particles* to be either bosons or fermions, corresponding respectively to symmetric and antisymmetric wavefunctions, is valid only for cases in which the particles move in three-or higher, dimensional space. The idea is that in a three-dimensional space a particle circulating around an identical one, can be continuously deformed into the identity. So, the wavefunction after the circulation should be equal to the original one. A full circulation is equivalent to two successive particle exchanges so that a single exchange should lead to a phase factor either 1 for bosons, or $-1$ for fermions. When we restrict the movement of the particles to two spatial dimensions there are more possibilities [13]. When a particle circulates around identical one in a two-dimensional space it is not possible to deform its movement continuously to the identity, as intersection with the other particle is not *topologically* allowed. In this case the wavefunction may acquire a phase $2\theta$ where $\theta$ will be different from $0$ or $\pi$. For two dimensional space the identical particles *may* follow different statistics referred to as 'anyons' [14].

What we can learn from the above discussions is that in a three dimensional space the total wavefunction should be of unique value. This 'conjecture' might explain the compatibility between our calculations for the AB phase [4] and that made by AB. I guess that such complementarity principle be valid also under different conditions [11,12].